# 1. Cosmological Parameters

Updated September 2019, by O. Lahav (University College London) and A.R. Liddle (Perimeter Institute for Theoretical Physics and Universidade de Lisboa).

## 1.1. Parametrizing the Universe

Rapid advances in observational cosmology have led to the establishment of a precision cosmological model, with many of the key cosmological parameters determined to one or two significant figure accuracy. Particularly prominent are measurements of cosmic microwave background (CMB) anisotropies, with the highest precision observations being those of the *Planck* Satellite [1,2] which supersede the landmark *WMAP* results [3,4]. However the most accurate model of the Universe requires consideration of a range of observations, with complementary probes providing consistency checks, lifting parameter degeneracies, and enabling the strongest constraints to be placed.

The term 'cosmological parameters' is forever increasing in its scope, and nowadays often includes the parameterization of some functions, as well as simple numbers describing properties of the Universe. The original usage referred to the parameters describing the global dynamics of the Universe, such as its expansion rate and curvature. Now we wish to know how the matter budget of the Universe is built up from its constituents: baryons, photons, neutrinos, dark matter, and dark energy. We also need to describe the nature of perturbations in the Universe, through global statistical descriptors such as the matter and radiation power spectra. There may be additional parameters describing the physical state of the Universe, such as the ionization fraction as a function of time during the era since recombination. Typical comparisons of cosmological models with observational data now feature between five and ten parameters.

### 1.1.1. *The global description of the Universe*:

Ordinarily, the Universe is taken to be a perturbed Robertson–Walker space-time, with dynamics governed by Einstein's equations. This is described in detail in the Big-Bang Cosmology chapter in this volume. Using the density parameters $\Omega_i$ for the various matter species and $\Omega_\Lambda$ for the cosmological constant, the Friedmann equation can be written

$$\sum_i \Omega_i + \Omega_\Lambda - 1 = \frac{k}{R^2 H^2}, \qquad (1.1)$$

where the sum is over all the different species of material in the Universe. This equation applies at any epoch, but later in this article we will use the symbols $\Omega_i$ and $\Omega_\Lambda$ to refer specifically to the present-epoch values.

The complete present-epoch state of the homogeneous Universe can be described by giving the current-epoch values of all the density parameters and the Hubble constant $h$ (the present-day Hubble parameter being written $H_0 = 100 h \, \text{km}\,\text{s}^{-1}\,\text{Mpc}^{-1}$). A typical collection would be baryons $\Omega_\text{b}$, photons $\Omega_\gamma$, neutrinos $\Omega_\nu$, and cold dark matter $\Omega_\text{c}$ (given charge neutrality, the electron density is guaranteed to be too small to be worth considering separately and is effectively included with the baryons). The spatial curvature can then be determined from the other parameters using Eq. (1.1). The total present matter density $\Omega_\text{m} = \Omega_\text{c} + \Omega_\text{b}$ may be used in place of the cold dark matter density $\Omega_\text{c}$.

These parameters also allow us to track the history of the Universe, at least back until an epoch where interactions allow interchanges between the densities of the different species; this is believed to have last happened at neutrino decoupling, shortly before Big-Bang Nucleosynthesis (BBN). To probe further back into the Universe's history requires assumptions about particle interactions, and perhaps about the nature of physical laws themselves.

The standard neutrino sector has three flavors. For neutrinos of mass in the range $5 \times 10^{-4}\,\mathrm{eV}$ to $1\,\mathrm{MeV}$, the density parameter in neutrinos is predicted to be

$$\Omega_\nu h^2 = \frac{\sum m_\nu}{93.14\,\mathrm{eV}}\,, \tag{1.2}$$

where the sum is over all families with mass in that range (higher masses need a more sophisticated calculation). We use units with $c = 1$ throughout. Results on atmospheric and Solar neutrino oscillations [5] imply non-zero mass-squared differences between the three neutrino flavors. These oscillation experiments cannot tell us the absolute neutrino masses, but within the simple assumption of a mass hierarchy suggest a lower limit of approximately $0.06\,\mathrm{eV}$ for the sum of the neutrino masses (see the Neutrino chapter).

Even a mass this small has a potentially observable effect on the formation of structure, as neutrino free-streaming damps the growth of perturbations. Analyses commonly now either assume a neutrino mass sum fixed at this lower limit, or allow the neutrino mass sum to be a variable parameter. To date there is no decisive evidence of any effects from either neutrino masses or an otherwise non-standard neutrino sector, and observations impose quite stringent limits; see the Neutrinos in Cosmology chapter. However, we note that the inclusion of the neutrino mass sum as a free parameter can affect the derived values of other cosmological parameters.

**1.1.2.** *Inflation and perturbations*:

A complete model of the Universe should include a description of deviations from homogeneity, at least in a statistical way. Indeed, some of the most powerful probes of the parameters described above come from the evolution of perturbations, so their study is naturally intertwined with the determination of cosmological parameters.

There are many different notations used to describe the perturbations, both in terms of the quantity used to and the definition of the statistical measure. We use the dimensionless power spectrum $\Delta^2$ as defined in the Big Bang Cosmology section (also denoted $\mathcal{P}$ in some of the literature). If the perturbations obey Gaussian statistics, the power spectrum provides a complete description of their properties.

From a theoretical perspective, a useful quantity to describe the perturbations is the curvature perturbation $\mathcal{R}$, which measures the spatial curvature of a comoving slicing of the space-time. A simple case is the Harrison–Zeldovich spectrum, which corresponds to a constant $\Delta^2_\mathcal{R}$. More generally, one can approximate the spectrum by a power law, writing

$$\Delta^2_\mathcal{R}(k) = \Delta^2_\mathcal{R}(k_*)\left[\frac{k}{k_*}\right]^{n_\mathrm{s}-1}, \tag{1.3}$$

where $n_{\rm s}$ is known as the spectral index, always defined so that $n_{\rm s} = 1$ for the Harrison–Zeldovich spectrum, and $k_*$ is an arbitrarily chosen scale. The initial spectrum, defined at some early epoch of the Universe's history, is usually taken to have a simple form such as this power law, and we will see that observations require $n_{\rm s}$ close to one. Subsequent evolution will modify the spectrum from its initial form.

The simplest mechanism for generating the observed perturbations is the inflationary cosmology, which posits a period of accelerated expansion in the Universe's early stages [6,7]. It is a useful working hypothesis that this is the sole mechanism for generating perturbations, and it may further be assumed to be the simplest class of inflationary model, where the dynamics are equivalent to that of a single scalar field $\phi$ with canonical kinetic energy slowly rolling on a potential $V(\phi)$. One may seek to verify that this simple picture can match observations and to determine the properties of $V(\phi)$ from the observational data. Alternatively, more complicated models, perhaps motivated by contemporary fundamental physics ideas, may be tested on a model-by-model basis (see more in the Inflation chapter in this volume).

Inflation generates perturbations through the amplification of quantum fluctuations, which are stretched to astrophysical scales by the rapid expansion. The simplest models generate two types, density perturbations that come from fluctuations in the scalar field and its corresponding scalar metric perturbation, and gravitational waves that are tensor metric fluctuations. The former experience gravitational instability and lead to structure formation, while the latter can influence the CMB anisotropies. Defining slow-roll parameters (with primes indicating derivatives with respect to the scalar field) as

$$\epsilon = \frac{m_{\rm Pl}^2}{16\pi}\left(\frac{V'}{V}\right)^2 \quad, \quad \eta = \frac{m_{\rm Pl}^2}{8\pi}\frac{V''}{V}, \tag{1.4}$$

which should satisfy $\epsilon, |\eta| \ll 1$, the spectra can be computed using the slow-roll approximation as

$$\Delta_{\cal R}^2(k) \simeq \left.\frac{8}{3m_{\rm Pl}^4}\frac{V}{\epsilon}\right|_{k=aH} \quad, \quad \Delta_{\rm t}^2(k) \simeq \left.\frac{128}{3m_{\rm Pl}^4}V\right|_{k=aH}. \tag{1.5}$$

In each case, the expressions on the right-hand side are to be evaluated when the scale $k$ is equal to the Hubble radius during inflation. The symbol '$\simeq$' here indicates use of the slow-roll approximation, which is expected to be accurate to a few percent or better.

From these expressions, we can compute the spectral indices [8]:

$$n_{\rm s} \simeq 1 - 6\epsilon + 2\eta \quad ; \quad n_{\rm t} \simeq -2\epsilon. \tag{1.6}$$

Another useful quantity is the ratio of the two spectra, defined by

$$r \equiv \frac{\Delta_{\rm t}^2(k_*)}{\Delta_{\cal R}^2(k_*)}. \tag{1.7}$$

We have

$$r \simeq 16\epsilon \simeq -8n_{\rm t}, \tag{1.8}$$

which is known as the consistency equation.

One could consider corrections to the power-law approximation, which we discuss later. However, for now we make the working assumption that the spectra can be approximated by such power laws. The consistency equation shows that $r$ and $n_t$ are not independent parameters, and so the simplest inflation models give initial conditions described by three parameters, usually taken as $\Delta_{\mathcal{R}}^2$, $n_s$, and $r$, all to be evaluated at some scale $k_*$, usually the 'statistical center' of the range explored by the data. Alternatively, one could use the parametrization $V$, $\epsilon$, and $\eta$, all evaluated at a point on the putative inflationary potential.

After the perturbations are created in the early Universe, they undergo a complex evolution up until the time they are observed in the present Universe. When the perturbations are small, this can be accurately followed using a linear theory numerical code such as CAMB or CLASS [9]. This works right up to the present for the CMB, but for density perturbations on small scales non-linear evolution is important and can be addressed by a variety of semi-analytical and numerical techniques. However the analysis is made, the outcome of the evolution is in principle determined by the cosmological model and by the parameters describing the initial perturbations, and hence can be used to determine them.

Of particular interest are CMB anisotropies. Both the total intensity and two independent polarization modes are predicted to have anisotropies. These can be described by the radiation angular power spectra $C_\ell$ as defined in the CMB article in this volume, and again provide a complete description if the density perturbations are Gaussian.

### 1.1.3. *The standard cosmological model*:

We now have most of the ingredients in place to describe the cosmological model. Beyond those of the previous subsections, we need a measure of the ionization state of the Universe. The Universe is known to be highly ionized at low redshifts (otherwise radiation from distant quasars would be heavily absorbed in the ultra-violet), and the ionized electrons can scatter microwave photons, altering the pattern of observed anisotropies. The most convenient parameter to describe this is the optical depth to scattering $\tau$ (*i.e.*, the probability that a given photon scatters once); in the approximation of instantaneous and complete reionization, this could equivalently be described by the redshift of reionization $z_{\mathrm{ion}}$.

As described in Sec. 1.4, models based on these parameters are able to give a good fit to the complete set of high-quality data available at present, and indeed some simplification is possible. Observations are consistent with spatial flatness, and the inflation models so far described automatically generate negligible spatial curvature, so we can set $k = 0$; the density parameters then must sum to unity, and so one of them can be eliminated. The neutrino energy density is often not taken as an independent parameter; provided that the neutrino sector has the standard interactions, the neutrino energy density, while relativistic, can be related to the photon density using thermal physics arguments, and a minimal assumption takes the neutrino mass sum to be that of the lowest mass solution to the neutrino oscillation constraints, namely 0.06 eV. In addition, there is no observational evidence for the existence of tensor perturbations (though the upper limits are fairly weak), and so $r$ could be set to zero. This leaves seven parameters, which is the smallest set that can usefully be compared to the present cosmological data. This model is referred to by various names, including $\Lambda$CDM, the concordance cosmology, and the standard cosmological model.

Of these parameters, only $\Omega_\gamma$ is accurately measured directly. The radiation density is dominated by the energy in the CMB, and the COBE satellite FIRAS experiment determined its temperature to be $T = 2.7255 \pm 0.0006\,\text{K}$ [10], [‡] corresponding to $\Omega_\gamma = 2.47 \times 10^{-5} h^{-2}$. It typically can be taken as fixed when fitting other data. Hence the minimum number of cosmological parameters varied in fits to data is six, though as described below there may additionally be many 'nuisance' parameters necessary to describe astrophysical processes influencing the data.

In addition to this minimal set, there is a range of other parameters that might prove important in future as the data-sets further improve, but for which there is so far no direct evidence, allowing them to be set to specific values for now. We discuss various speculative options in the next section. For completeness at this point, we mention one other interesting quantity, the helium fraction, which is a non-zero parameter that can affect the CMB anisotropies at a subtle level. It is usually fixed in microwave anisotropy studies, but the data are approaching a level where allowing its variation may become mandatory.

Most attention to date has been on parameter estimation, where a set of parameters is chosen by hand and the aim is to constrain them. Interest has been growing towards the higher-level inference problem of model selection, which compares different choices of parameter sets. Bayesian inference offers an attractive framework for cosmological model selection, setting a tension between model predictiveness and ability to fit the data [11].

**1.1.4.** *Derived parameters*:

The parameter list of the previous subsection is sufficient to give a complete description of cosmological models that agree with observational data. However, it is not a unique parameterization, and one could instead use parameters derived from that basic set. Parameters that can be obtained from the set given above include the age of the Universe, the present horizon distance, the present neutrino background temperature, the epoch of matter–radiation equality, the epochs of recombination and decoupling, the epoch of transition to an accelerating Universe, the baryon-to-photon ratio, and the baryon-to-dark-matter density ratio. In addition, the physical densities of the matter components, $\Omega_i h^2$, are often more useful than the density parameters. The density perturbation amplitude can be specified in many different ways other than the large-scale primordial amplitude, for instance, in terms of its effect on the CMB, or by specifying a short-scale quantity, a common choice being the present linear-theory mass dispersion on a scale of $8\,h^{-1}$Mpc, known as $\sigma_8$.

---

[‡] Unless stated otherwise, all quoted uncertainties in this article are $1\sigma/68\%$ confidence and all upper limits are 95% confidence. Cosmological parameters sometimes have significantly non-Gaussian uncertainties. Throughout we have rounded central values, and especially uncertainties, from original sources, in cases where they appear to be given to excessive precision.

Different types of observation are sensitive to different subsets of the full cosmological parameter set, and some are more naturally interpreted in terms of some of the derived parameters of this subsection than on the original base parameter set. In particular, most types of observation feature degeneracies whereby they are unable to separate the effects of simultaneously varying specific combinations of several of the base parameters.

## 1.2. Extensions to the standard model

At present, there is no positive evidence in favor of extensions of the standard model. These are becoming increasingly constrained by the data, though there always remains the possibility of trace effects at a level below present observational capability.

### 1.2.1. *More general perturbations*:

The standard cosmology assumes adiabatic, Gaussian perturbations. Adiabaticity means that all types of material in the Universe share a common perturbation, so that if the space-time is foliated by constant-density hypersurfaces, then all fluids and fields are homogeneous on those slices, with the perturbations completely described by the variation of the spatial curvature of the slices. Gaussianity means that the initial perturbations obey Gaussian statistics, with the amplitudes of waves of different wavenumbers being randomly drawn from a Gaussian distribution of width given by the power spectrum. Note that gravitational instability generates non-Gaussianity; in this context, Gaussianity refers to a property of the initial perturbations, before they evolve.

The simplest inflation models, based on one dynamical field, predict adiabatic perturbations and a level of non-Gaussianity that is too small to be detected by any experiment so far conceived. For present data, the primordial spectra are usually assumed to be power laws.

#### 1.2.1.1. *Non-power-law spectra:*

For typical inflation models, it is an approximation to take the spectra as power laws, albeit usually a good one. As data quality improves, one might expect this approximation to come under pressure, requiring a more accurate description of the initial spectra, particularly for the density perturbations. In general, one can expand $\ln \Delta_{\mathcal{R}}^2$ as

$$\ln \Delta_{\mathcal{R}}^2 (k) = \ln \Delta_{\mathcal{R}}^2 (k_*) + \left( n_{\mathrm{s},*} - 1 \right) \ln \frac{k}{k_*} + \frac{1}{2} \left. \frac{dn_{\mathrm{s}}}{d \ln k} \right|_* \ln^2 \frac{k}{k_*} + \cdots , \quad (1.9)$$

where the coefficients are all evaluated at some scale $k_*$. The term $dn_{\mathrm{s}}/d\ln k|_*$ is often called the running of the spectral index [12]. Once non-power-law spectra are allowed, it is necessary to specify the scale $k_*$ at which the spectral index is defined.

**1.2.1.2.** *Isocurvature perturbations:*

An isocurvature perturbation is one that leaves the total density unperturbed, while perturbing the relative amounts of different materials. If the Universe contains $N$ fluids, there is one growing adiabatic mode and $N-1$ growing isocurvature modes (for reviews see Ref. 7 and Ref. 13). These can be excited, for example, in inflationary models where there are two or more fields that acquire dynamically-important perturbations. If one field decays to form normal matter, while the second survives to become the dark matter, this will generate a cold dark matter isocurvature perturbation.

In general, there are also correlations between the different modes, and so the full set of perturbations is described by a matrix giving the spectra and their correlations. Constraining such a general construct is challenging, though constraints on individual modes are beginning to become meaningful, with no evidence that any other than the adiabatic mode must be non-zero.

**1.2.1.3.** *Seeded perturbations:*

An alternative to laying down perturbations at very early epochs is that they are seeded throughout cosmic history, for instance by topological defects such as cosmic strings. It has long been excluded that these are the sole original of structure, but they could contribute part of the perturbation signal, current limits being just a few percent [14]. In particular, cosmic defects formed in a phase transition ending inflation is a plausible scenario for such a contribution.

**1.2.1.4.** *Non-Gaussianity:*

Multi-field inflation models can also generate primordial non-Gaussianity (reviewed, *e.g.*, in Ref. 7). The extra fields can either be in the same sector of the underlying theory as the inflaton, or completely separate, an interesting example of the latter being the curvaton model [15]. Current upper limits on non-Gaussianity are becoming stringent, but there remains strong motivation to push down those limits and perhaps reveal trace non-Gaussianity in the data. If non-Gaussianity is observed, its nature may favor an inflationary origin, or a different one such as topological defects.

**1.2.2.** *Dark matter properties*:

Dark matter properties are discussed in the Dark Matter chapter in this volume. The simplest assumption concerning the dark matter is that it has no significant interactions with other matter, and that its particles have a negligible velocity as far as structure formation is concerned. Such dark matter is described as 'cold,' and candidates include the lightest supersymmetric particle, the axion, and primordial black holes. As far as astrophysicists are concerned, a complete specification of the relevant cold dark matter properties is given by the density parameter $\Omega_c$, though those seeking to detect it directly need also to know its interaction properties.

Cold dark matter is the standard assumption and gives an excellent fit to observations, except possibly on the shortest scales where there remains some controversy concerning the structure of dwarf galaxies and possible substructure in galaxy halos. It has long been excluded for all the dark matter to have a large velocity dispersion, so-called 'hot' dark matter, as it does not permit galaxies to form; for thermal relics the mass must be above about 1 keV to satisfy this constraint, though relics produced non-thermally, such as the axion, need not obey this limit. However, in future further parameters might need to be introduced to describe dark matter properties relevant to astrophysical observations. Suggestions that have been made include a modest velocity dispersion (warm dark matter) and dark matter self-interactions. There remains the possibility that the dark matter is comprized of two separate components, *e.g.*, a cold one and a hot one, an example being if massive neutrinos have a non-negligible effect.

### 1.2.3.  *Relativistic species*:

The number of relativistic species in the young Universe (omitting photons) is denoted $N_{\rm eff}$. In the standard cosmological model only the three neutrino species contribute, and its baseline value is assumed fixed at 3.045 (the small shift from 3 is because of a slight predicted deviation from a thermal distribution [16]) . However other species could contribute, for example an extra neutrino, possibly of sterile type, or massless Goldstone bosons or other scalars. It is hence interesting to study the effect of allowing this parameter to vary, and indeed although 3.045 is consistent with the data, most analyses currently suggest a somewhat higher value (*e.g.*, Ref. 17).

### 1.2.4.  *Dark energy*:

While the standard cosmological model given above features a cosmological constant, in order to explain observations indicating that the Universe is presently accelerating, further possibilities exist under the general headings of 'dark energy' and 'modified gravity'. These topics are described in detail in the Dark Energy chapter in this volume. This article focuses on the case of the cosmological constant, since this simple model is a good match to existing data. We note that more general treatments of dark energy/modified gravity will lead to weaker constraints on other parameters.

### 1.2.5.  *Complex ionization history*:

The full ionization history of the Universe is given by the ionization fraction as a function of redshift $z$. The simplest scenario takes the ionization to have the small residual value left after recombination up to some redshift $z_{\rm ion}$, at which point the Universe instantaneously reionizes completely. Then there is a one-to-one correspondence between $\tau$ and $z_{\rm ion}$ (that relation, however, also depending on other cosmological parameters). An accurate treatment of this process will track separate histories for hydrogen and helium. While currently rapid ionization appears to be a good approximation, as data improve a more complex ionization history may need to be considered.

### 1.2.6.  *Varying 'constants'*:

Variation of the fundamental constants of Nature over cosmological times is another possible enhancement of the standard cosmology. There is a long history of study of variation of the gravitational constant $G_{\rm N}$, and more recently attention has been drawn to the possibility of small fractional variations in the fine-structure constant. There is presently no observational evidence for the former, which is tightly constrained by a variety of measurements. Evidence for the latter has been claimed from studies of spectral line shifts in quasar spectra at redshift $z \approx 2$ [18], but this is presently controversial and in need of further observational study.

**1.2.7.** *Cosmic topology*:

The usual hypothesis is that the Universe has the simplest topology consistent with its geometry, for example that a flat universe extends forever. Observations cannot tell us whether that is true, but they can test the possibility of a non-trivial topology on scales up to roughly the present Hubble scale. Extra parameters would be needed to specify both the type and scale of the topology; for example, a cuboidal topology would need specification of the three principal axis lengths and orientation. At present, there is no evidence for non-trivial cosmic topology [19].

## 1.3. Cosmological Probes

The goal of the observational cosmologist is to utilize astronomical information to derive cosmological parameters. The transformation from the observables to the parameters usually involves many assumptions about the nature of the data, as well as of the dark sector. Below we outline the physical processes involved in each of the major probes, and the main recent results. The first two subsections concern probes of the homogeneous Universe, while the remainder consider constraints from perturbations.

In addition to statistical uncertainties we note three sources of systematic uncertainties that will apply to the cosmological parameters of interest: (i) due to the assumptions on the cosmological model and its priors (*i.e.*, the number of assumed cosmological parameters and their allowed range); (ii) due to the uncertainty in the astrophysics of the objects (*e.g.*, light-curve fitting for supernovae or the mass–temperature relation of galaxy clusters); and (iii) due to instrumental and observational limitations (*e.g.*, the effect of 'seeing' on weak gravitational lensing measurements, or beam shape on CMB anisotropy measurements).

These systematics, the last two of which appear as 'nuisance parameters', pose a challenging problem to the statistical analysis. We attempt a statistical fit to the whole Universe with 6 to 12 parameters, but we might need to include hundreds of nuisance parameters, some of them highly correlated with the cosmological parameters of interest (for example time-dependent galaxy biasing could mimic the growth of mass fluctuations). Fortunately, there is some astrophysical prior knowledge on these effects, and a small number of physically-motivated free parameters would ideally be preferred in the cosmological parameter analysis.

**1.3.1.** *Measures of the Hubble constant*:

In 1929, Edwin Hubble discovered the law of expansion of the Universe by measuring distances to nearby galaxies. The slope of the relation between the distance and recession velocity is defined to be the present-epoch Hubble constant, $H_0$. Astronomers argued for decades about the systematic uncertainties in various methods and derived values over the wide range $40 \, \mathrm{km \, s^{-1} \, Mpc^{-1}} \lesssim H_0 \lesssim 100 \, \mathrm{km \, s^{-1} \, Mpc^{-1}}$.

One of the most reliable results on the Hubble constant came from the Hubble Space Telescope (HST) Key Project [20]. This study used the empirical period–luminosity relation for Cepheid variable stars, and calibrated a number of secondary distance indicators—Type Ia Supernovae (SNe Ia), the Tully–Fisher relation, surface-brightness fluctuations, and Type II Supernovae. This approach was further extended, based on HST observations of 70 long-period Cepheids in the Large Magellanic Cloud, combined with Milky Way parallaxes and masers in NGC4258, to yield $H_0 = 74.0 \pm 1.4 \, \text{km s}^{-1} \, \text{Mpc}^{-1}$ [21] (the SH0ES project). The major sources of uncertainty in this result are thought to be due to the heavy element abundance of the Cepheids and the distance to the fiducial nearby galaxy, the Large Magellanic Cloud, relative to which all Cepheid distances are measured.

Three other methods have been used recently. One is a calibration of the tip of the red-giant branch applied to Type Ia supernovae, the Carnegie–Chicago Hubble Programme (CCHP) finding $H_0 = 69.8 \pm 0.8 \, (\text{stat.}) \pm 1.7 \, (\text{sys.}) \, \text{km s}^{-1} \, \text{Mpc}^{-1}$ [22]. The second uses the method of time delay in six gravitationally-lensed quasars, with the result $H_0 = 73.3^{+1.7}_{-1.8} \, \text{km s}^{-1} \, \text{Mpc}^{-1}$ [23] (H0LiCOW). A third method that came to fruition recently is based on gravitational waves; the 'bright standard siren' applied to the binary neutron star GW170817 and the 'dark standard siren' implemented on the binary black hole GW170814 yield $H_0 = 70^{+12}_{-8} \, \text{km s}^{-1} \, \text{Mpc}^{-1}$ [24] and $H_0 = 75^{+40}_{-32} \, \text{km s}^{-1} \, \text{Mpc}^{-1}$ [25] respectively. With many more gravitational-wave events the future uncertainties on $H_0$ from standard sirens will get smaller.

The determination of $H_0$ by the *Planck* Collaboration [2] gives a lower value, $H_0 = 67.4 \pm 0.5 \, \text{km s}^{-1} \, \text{Mpc}^{-1}$. As discussed in their paper, there is strong degeneracy of $H_0$ with other parameters, *e.g.*, $\Omega_m$ and the neutrino mass. It is worth noting that using the 'inverse distance ladder' method gives a result $H_0 = 67.8 \pm 1.3 \, \text{km s}^{-1} \, \text{Mpc}^{-1}$ [26], close to the *Planck* result. The inverse distance ladder relies on absolute-distance measurements from baryon acoustic oscillations (BAOs) to calibrate the intrinsic magnitude of the SNe Ia (rather than by nearby Cepheids and parallax). This measurement was derived from 207 spectroscopically-confirmed Type Ia supernovae from the Dark Energy Survey (DES), an additional 122 low-redshift SNe Ia, and measurements of BAOs. A combination of DES Y1 clustering and weak lensing with BAO and BBN (assuming ΛCDM) gives $H_0 = 67.4^{+1.1}_{-1.2} \, \text{km s}^{-1} \, \text{Mpc}^{-1}$ [27].

The tension between the $H_0$ values from *Planck* and the traditional cosmic distance ladder methods is of great interest and under investigation. For example, the SH0ES and H0LiCOW+SH0ES results deviate from *Planck* by $4.4\sigma$ and $5.3\sigma$ respectively, while the TRGB and standard-siren results lie between the *Planck* and cosmic ladder $H_0$ values. There is possibly a trend for higher $H_0$ derived from the nearby Universe and a lower $H_0$ from the early Universe, which has led some researchers to propose a time-variation of the dark energy component or other exotic scenarios. Ongoing studies are addressing the question of whether the Hubble tension is due to systematics in at least one of the probes, or a signature of new physics.

Figure 1.1 shows a selection of recent $H_0$ values, adapted from Ref. 28 which provides a very useful summary of the current status of the Hubble constant tension.

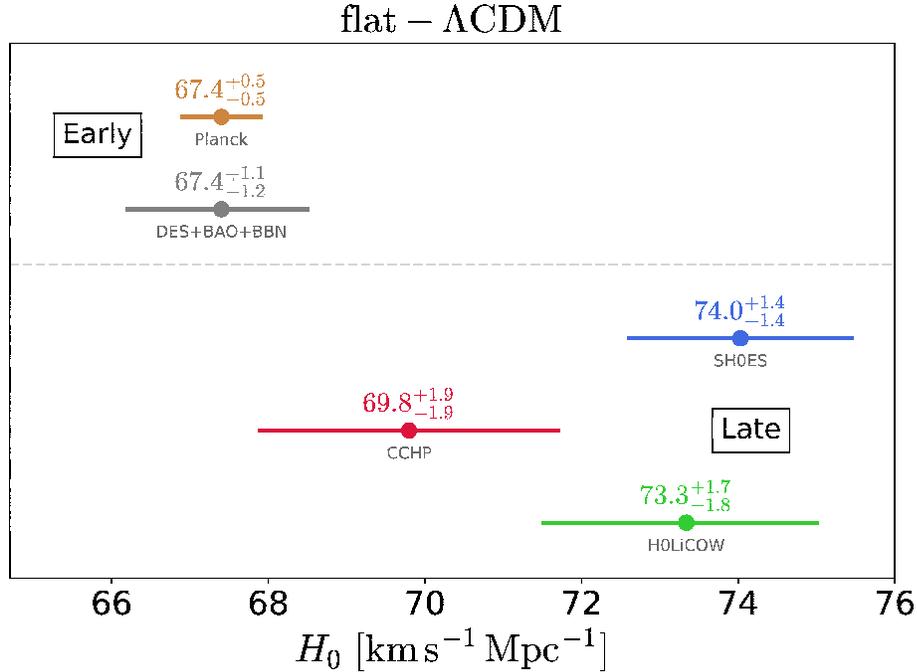

**Figure 1.1:** A selection of recent $H_0$ measurements from the various projects as described in the text, divided into early and late Universe probes. The standard-siren determinations are omitted as they are too wide for the plot. Figure courtesy of Vivien Bonvin and Martin Millon, adapted from Ref. 28.

### 1.3.2. *Supernovae as cosmological probes*:

Empirically, the peak luminosity of SNe Ia can be used as an efficient distance indicator (*e.g.*, Ref. 29), thus allowing cosmology to be constrained via the distance–redshift relation. The favorite theoretical explanation for SNe Ia is the thermonuclear disruption of carbon–oxygen white dwarfs. Although not perfect 'standard candles', it has been demonstrated that by correcting for a relation between the light-curve shape, color, and luminosity at maximum brightness, the dispersion of the measured luminosities can be greatly reduced. There are several possible systematic effects that may affect the accuracy of the use of SNe Ia as distance indicators, *e.g.*, evolution with redshift and interstellar extinction in the host galaxy and in the Milky Way.

Two major studies, the Supernova Cosmology Project and the High-$z$ Supernova Search Team, found evidence for an accelerating Universe [30], interpreted as due to a cosmological constant or a dark energy component. When combined with the CMB data (which indicate near flatness, *i.e.*, $\Omega_m + \Omega_\Lambda \simeq 1$), the best-fit values were $\Omega_m \approx 0.3$ and $\Omega_\Lambda \approx 0.7$. Most results in the literature are consistent with the $w = -1$ cosmological constant case. One study [31] deduced, from a sample of 740 spectroscopically-confirmed SNe Ia, that $\Omega_m = 0.295 \pm 0.034$ (stat+sym) for an assumed flat $\Lambda$CDM model. An analysis of a sample of spectroscopically-confirmed 207 DES SNe Ia combined with 122 low-redshift SNe [32] yielded $\Omega_m = 0.331 \pm 0.038$ for an assumed flat $\Lambda$CDM model. In combination with the CMB, for a flat $w$CDM these data give $w = -0.978 \pm 0.059$ and $\Omega_{\rm m} = 0.321 \pm 0.018$, consistent with results from the JLA and Pantheon SNe Ia samples. Future experiments will refine constraints on the cosmic equation of state $w(z)$.

### 1.3.3.   *Cosmic microwave background*:

The physics of the CMB is described in detail in the CMB chapter in this volume. Before recombination, the baryons and photons are tightly coupled, and the perturbations oscillate in the potential wells generated primarily by the dark matter perturbations. After decoupling, the baryons are free to collapse into those potential wells. The CMB carries a record of conditions at the time of last scattering, often called primary anisotropies. In addition, it is affected by various processes as it propagates towards us, including the effect of a time-varying gravitational potential (the integrated Sachs–Wolfe effect), gravitational lensing, and scattering from ionized gas at low redshift.

The primary anisotropies, the integrated Sachs–Wolfe effect, and the scattering from a homogeneous distribution of ionized gas, can all be calculated using linear perturbation theory. Available codes include CAMB and CLASS [9], the former widely used embedded within the analysis package CosmoMC [33] and in higher-level analysis packages such as CosmoSIS [34] and CosmoLike [35]. Gravitational lensing is also calculated in these codes. Secondary effects, such as inhomogeneities in the reionization process, and scattering from gravitationally-collapsed gas (the Sunyaev–Zeldovich or SZ effect), require more complicated, and more uncertain, calculations.

The upshot is that the detailed pattern of anisotropies depends on all of the cosmological parameters. In a typical cosmology, the anisotropy power spectrum [usually plotted as $\ell(\ell+1)C_\ell$] features a flat plateau at large angular scales (small $\ell$), followed by a series of oscillatory features at higher angular scales, the first and most prominent being at around one degree ($\ell \simeq 200$). These features, known as acoustic peaks, represent the oscillations of the photon–baryon fluid around the time of decoupling. Some features can be closely related to specific parameters—for instance, the location in multipole space of the set of peaks probes the spatial geometry, while the relative heights of the peaks probe the baryon density—but many other parameters combine to determine the overall shape.

The 2018 data release from the *Planck* satellite [1] gives the most powerful results to date on the spectrum of CMB temperature anisotropies, with a precision determination of the temperature power spectrum to beyond $\ell = 2000$. The Atacama Cosmology Telescope (ACT) and South Pole Telescope (SPT) experiments extend these results to higher angular resolution, though without full-sky coverage. *Planck* and the polarisation-sensitive versions of ACT and SPT give the state of the art in measuring the spectrum of $E$-polarization anisotropies and the correlation spectrum between temperature and polarization. These are consistent with models based on the parameters we have described, and provide accurate determinations of many of those parameters [2]. Primordial $B$-mode polarization has not been detected (although the gravitational lensing effect on $B$ modes has been measured).

The data provide an exquisite measurement of the location of the set of acoustic peaks, determining the angular-diameter distance of the last-scattering surface. In combination with other data this strongly constrains the spatial geometry, in a manner consistent with spatial flatness and excluding significantly-curved Universes. CMB data give a precision measurement of the age of the Universe. The CMB also gives a baryon density consistent with, and at higher precision than, that coming from BBN. It affirms the need for both dark matter and dark energy. It shows no evidence for dynamics of the dark energy, being consistent with a pure cosmological constant ($w = -1$). The density perturbations are consistent with a power-law primordial spectrum, and there is no indication yet of tensor perturbations. The current best-fit for the reionization optical depth from CMB data, $\tau = 0.054$, is in line with models of how early structure formation induces reionization.

*Planck* has also made the first all-sky map of the CMB lensing field, which probes the entire matter distribution in the Universe and adds some additional constraining power to the CMB-only data-sets. These measurements are compatible with the expected effect in the standard cosmology.

**1.3.4.** *Galaxy clustering*:

The power spectrum of density perturbations is affected by the nature of the dark matter. Within the ΛCDM model, the power spectrum shape depends primarily on the primordial power spectrum and on the combination $\Omega_\mathrm{m} h$, which determines the horizon scale at matter–radiation equality, with a subdominant dependence on the baryon density. The matter distribution is most easily probed by observing the galaxy distribution, but this must be done with care since the galaxies do not perfectly trace the dark matter distribution. Rather, they are a 'biased' tracer of the dark matter [36]. The need to allow for such bias is emphasized by the observation that different types of galaxies show bias with respect to each other. In particular, scale-dependent and stochastic biasing may introduce a systematic effect on the determination of cosmological parameters from redshift surveys [37]. Prior knowledge from simulations of galaxy formation or from gravitational lensing data could help to quantify biasing. Furthermore, the observed 3D galaxy distribution is in redshift space, *i.e.*, the observed redshift is the sum of the Hubble expansion and the line-of-sight peculiar velocity, leading to linear and non-linear dynamical effects that also depend on the cosmological parameters. On the largest length scales, the galaxies are expected to trace the location of the dark matter, except for a constant multiplier $b$ to the power spectrum, known as the linear bias parameter. On scales smaller than 20 Mpc or so, the clustering pattern is 'squashed' in the radial direction due to coherent infall, which depends approximately on the parameter $\beta \equiv \Omega_\mathrm{m}^{0.6}/b$ (on these shorter scales, more complicated forms of biasing are not excluded by the data). On scales of a few Mpc, there is an effect of elongation along the line of sight (colloquially known as the 'finger of God' effect) that depends on the galaxy velocity dispersion.

**1.3.4.1.** *Baryon acoustic oscillations:*

The power spectra of the 2-degree Field (2dF) Galaxy Redshift Survey and the Sloan Digital Sky Survey (SDSS) are well fit by a ΛCDM model and both surveys showed first evidence for baryon acoustic oscillations (BAOs) [38,39]. The Baryon Oscillation Spectroscopic Survey (BOSS) of luminous red galaxies (LRGs) in the SDSS (DR 12) found, using a sample of 1.2 million galaxies, consistency with $w = -1.01 \pm 0.06$ [40] when combined with *Planck* 2015. Similar results for $w$ were obtained by the WiggleZ survey [41].

**1.3.4.2.**   *Redshift distortion:*

There is renewed interest in the 'redshift distortion' effect. This distortion depends on cosmological parameters [42] via the perturbation growth rate in linear theory $f(z) = d\ln\delta/d\ln a \approx \Omega^\gamma(z)$, where $\gamma \simeq 0.55$ for the $\Lambda$CDM model and may be different for modified gravity models. By measuring $f(z)$ it is feasible to constrain $\gamma$ and rule out certain modified gravity models [43,44]. We note the degeneracy of the redshift-distortion pattern and the geometric distortion (the so-called Alcock–Paczynski effect [45]), *e.g.*, as illustrated by the WiggleZ survey [46] and the BOSS Survey [47].

**1.3.4.3.**   *Limits on neutrino mass from galaxy surveys and other probes:*

Large-scale structure data place constraints on $\Omega_\nu$ due to the neutrino free-streaming effect [48]. Presently there is no clear detection, and upper limits on neutrino mass are commonly estimated by comparing the observed galaxy power spectrum with a four-component model of baryons, cold dark matter, a cosmological constant, and massive neutrinos. Such analyses also assume that the primordial power spectrum is adiabatic, scale-invariant, and Gaussian. Potential systematic effects include biasing of the galaxy distribution and non-linearities of the power spectrum. An upper limit can also be derived from CMB anisotropies alone, while combination with additional cosmological data-sets can improve the results.

The most recent results on neutrino mass upper limits and other neutrino properties are summarised in the Neutrinos in Cosmology chapter in this volume. While the latest cosmological data do not yet constrain the sum of neutrino masses to below 0.2 eV, since the lower limit on this sum from oscillation experiments is 0.06 eV it is expected that future cosmological surveys will soon detect effects from the neutrino mass. Also, current cosmological datasets are in good agreement with the standard value for the effective number of neutrino species $N_{\text{eff}} = 3.045$.

**1.3.5.**   *Clustering in the inter-galactic medium*:

It is commonly assumed, based on hydrodynamic simulations, that the neutral hydrogen in the inter-galactic medium (IGM) can be related to the underlying mass distribution. It is then possible to estimate the matter power spectrum on scales of a few megaparsecs from the absorption observed in quasar spectra, the so-called Lyman-$\alpha$ forest. The usual procedure is to measure the power spectrum of the transmitted flux, and then to infer the mass power spectrum. Photo-ionization heating by the ultraviolet background radiation and adiabatic cooling by the expansion of the Universe combine to give a simple power-law relation between the gas temperature and the baryon density. It also follows that there is a power-law relation between the optical depth $\tau$ and $\rho_{\text{b}}$. Therefore, the observed flux $F = \exp(-\tau)$ is strongly correlated with $\rho_{\text{b}}$, which itself traces the mass density. The matter and flux power spectra can be related by a biasing function that is calibrated from simulations.

A study of 266,590 quasars in the range $1.77 < z < 3$ from SDSS was used to measure the BAO scale from the 3D correlation of Lyman-$\alpha$ and quasars [49]. Combined with the Lyman-$\alpha$ auto-correlation measurement presented in a companion paper [50] the BAO measurements at $z = 2.34$ are within $1.7\sigma$ of the *Planck* 2018 $\Lambda$CDM model. The Lyman-$\alpha$ flux power spectrum has also been used to constrain the nature of dark matter, for example limiting the amount of warm dark matter [51].

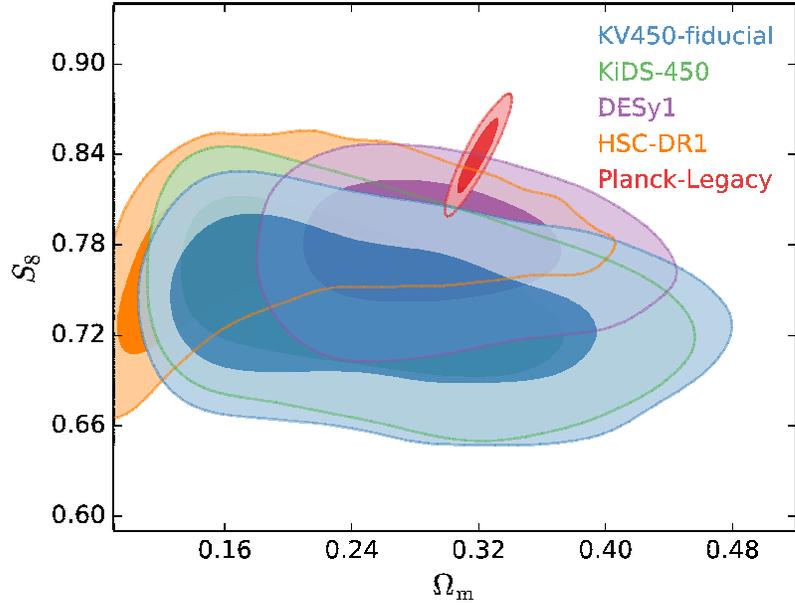

**Figure 1.2:** Marginalised posterior contours (inner 68% confidence level, outer 95% confidence level) in the $\Omega_{\mathrm{m}}$–$S_8$ plane. Shown are the optical-only KiDS-450 analysis (green; Ref. 53), the fiducial KiDS+VISTA-450 setup (blue; Ref. 53), DES Year 1 using cosmic shear only (purple; Ref. 54), HSC-DR1 cosmic shear (orange; Ref. 55) and the *Planck* Legacy analysis (red; *Planck* Collaboration [2] using TT+TE+EE+lowE). Figure from Ref. 53.

### 1.3.6. *Weak gravitational lensing*:

Images of background galaxies are distorted by the gravitational effect of mass variations along the line of sight. Deep gravitational potential wells, such as galaxy clusters, generate 'strong lensing' leading to arcs, arclets, and multiple images, while more moderate perturbations give rise to 'weak lensing'. Weak lensing is now widely used to measure the mass power spectrum in selected regions of the sky (see Ref. 52 for reviews). Since the signal is weak, the image of deformed galaxy shapes (the 'shear map') must be analyzed statistically to measure the power spectrum, higher moments, and cosmological parameters. There are various systematic effects in the interpretation of weak lensing, *e.g.*, due to atmospheric distortions during observations, the redshift distribution of the background galaxies (usually depending on the accuracy of photometric redshifts), the intrinsic correlation of galaxy shapes, and non-linear modeling uncertainties.

As one example, the 'Kilo-Degree Survey' (KiDS), combined with the VISTA VIKING survey, used weak-lensing measurements over 450 deg$^2$ to constrain the clumpiness parameter $S_8 \equiv \sigma_8(\Omega_{\mathrm{m}}/0.3)^{0.5} = 0.737^{+0.040}_{-0.036}$ [53]. This is lower by $2.3\sigma$ than $S_8$ derived from *Planck*. Figure 1.2 (which is Figure 4 from Ref. 53) shows the $\Omega_{\mathrm{m}}$–$S_8$ constraints derived from weak lensing of KiDS, DES, and HPC versus the CMB constraint from *Planck*. Variations in $S_8$ among the weak-lensing surveys are mainly due to difference in the procedures for photometric redshift determinations. Results from weak lensing from DES, combined with other probes, are shown in the next section.

**1.3.7.  *Other probes*:**

Other probes that have been used to constrain cosmological parameters, but that are not presently competitive in terms of accuracy, are the integrated Sachs–Wolfe effect [56,57], the number density or composition of galaxy clusters [58], and galaxy peculiar velocities, which probe the mass fluctuations in the local Universe [59].

## 1.4.  Bringing probes together

Although it contains two ingredients—dark matter and dark energy—which have not yet been verified by laboratory experiments, the ΛCDM model is almost universally accepted by cosmologists as the best description of the present data. The approximate values of some of the key parameters are $\Omega_b \approx 0.05$, $\Omega_c \approx 0.25$, $\Omega_\Lambda \approx 0.70$, and a Hubble constant $h \approx 0.70$. The spatial geometry is very close to flat (and usually assumed to be precisely flat), and the initial perturbations Gaussian, adiabatic, and nearly scale-invariant.

The most powerful data source is the CMB, which on its own supports all these main tenets. Values for some parameters, as given in Ref. 2, are reproduced in Table 1.1. These particular results presume a flat Universe. The constraints are somewhat strengthened by adding additional data-sets, BAO being shown in the Table as an example, though most of the constraining power resides in the CMB data. Similar constraints at lower precision were previously obtained by the *WMAP* collaboration.

If the assumption of spatial flatness is lifted, it turns out that the primary CMB on its own constrains the spatial curvature fairly weakly, due to a parameter degeneracy in the angular-diameter distance. However, inclusion of other data readily removes this degeneracy. Simply adding the *Planck* lensing measurement, and with the assumption that the dark energy is a cosmological constant, yields a 68% confidence constraint on $\Omega_{\rm tot} \equiv \sum \Omega_i + \Omega_\Lambda = 1.011 \pm 0.006$ and further adding BAO makes it $0.9993 \pm 0.0019$ [2]. Results of this type are normally taken as justifying the restriction to flat cosmologies.

One derived parameter that is very robust is the age of the Universe, since there is a useful coincidence that for a flat Universe the position of the first peak is strongly correlated with the age. The CMB data give $13.797 \pm 0.023$ Gyr (assuming flatness). This is in good agreement with the ages of the oldest globular clusters and with radioactive dating.

The baryon density $\Omega_b$ is now measured with high accuracy from CMB data alone, and is consistent with and much more precise than the determination from BBN. The value quoted in the Big-Bang Nucleosynthesis chapter in this volume is $0.021 \leq \Omega_b h^2 \leq 0.024$ (95% confidence).

While $\Omega_\Lambda$ is measured to be non-zero with very high confidence, there is no evidence of evolution of the dark energy density. As described in the Dark Energy chapter in this volume, from a combination of CMB, weak gravitational lensing, SN, and BAO measurements, assuming a flat universe, Ref. 2 found $w = -1.028 \pm 0.031$, consistent with the cosmological constant case $w = -1$. Allowing more complicated forms of dark energy weakens the limits.

**Table 1.1:** Parameter constraints reproduced from Ref. 2 (Table 2, column 5), with some additional rounding. Both columns assume the ΛCDM cosmology with a power-law initial spectrum, no tensors, spatial flatness, a cosmological constant as dark energy, and the sum of neutrino masses fixed to 0.06 eV. Above the line are the six parameter combinations actually fit to the data ($\theta_{\rm MC}$ is a measure of the sound horizon at last scattering); those below the line are derived from these. The first column uses *Planck* primary CMB data plus the *Planck* measurement of CMB lensing. This column gives our present recommended values. The second column adds in data from a compilation of BAO measurements described in Ref. 2. The perturbation amplitude $\Delta_{\mathcal{R}}^2$ (denoted $A_{\rm s}$ in the original paper) is specified at the scale $0.05\,{\rm Mpc}^{-1}$. Uncertainties are shown at 68% confidence.

| | *Planck* TT,TE,EE+lowE+lensing | +BAO |
|---|---|---|
| $\Omega_{\rm b}h^2$ | $0.02237 \pm 0.00015$ | $0.02242 \pm 0.00014$ |
| $\Omega_{\rm c}h^2$ | $0.1200 \pm 0.0012$ | $0.1193 \pm 0.0009$ |
| $100\,\theta_{\rm MC}$ | $1.0409 \pm 0.0003$ | $1.0410 \pm 0.0003$ |
| $n_{\rm s}$ | $0.965 \pm 0.004$ | $0.966 \pm 0.004$ |
| $\tau$ | $0.054 \pm 0.007$ | $0.056 \pm 0.007$ |
| $\ln(10^{10}\Delta_{\mathcal{R}}^2)$ | $3.044 \pm 0.014$ | $3.047 \pm 0.014$ |
| $h$ | $0.674 \pm 0.005$ | $0.677 \pm 0.004$ |
| $\sigma_8$ | $0.811 \pm 0.006$ | $0.810 \pm 0.006$ |
| $\Omega_{\rm m}$ | $0.315 \pm 0.007$ | $0.311 \pm 0.006$ |
| $\Omega_\Lambda$ | $0.685 \pm 0.007$ | $0.689 \pm 0.006$ |

The data provide strong support for the main predictions of the simplest inflation models: spatial flatness and adiabatic, Gaussian, nearly scale-invariant density perturbations. But it is disappointing that there is no sign of primordial gravitational waves, with a 95% confidence upper limit from combining *Planck* with BICEP2/Keck Array BK15 data of $r < 0.06$ at the scale $0.002\,{\rm Mpc}^{-1}$ [61] (weakening somewhat if running is allowed). The spectral index is clearly required to be less than one by current data, though the strength of that conclusion can weaken if additional parameters are included in the model fits.

Tests have been made for various types of non-Gaussianity, a particular example being a parameter $f_{\rm NL}$ that measures a quadratic contribution to the perturbations. Various non-Gaussian shapes are possible (see Ref. 62 for details), and current constraints on the popular 'local', 'equilateral', and 'orthogonal' types (combining temperature and polarization data) are $f_{\rm NL}^{\rm local} = -1 \pm 5$, $f_{\rm NL}^{\rm equil} = -26 \pm 47$, and $f_{\rm NL}^{\rm ortho} = -38 \pm 24$ respectively (these look weak, but prominent non-Gaussianity requires the product $f_{\rm NL}\Delta_{\mathcal{R}}$ to be large, and $\Delta_{\mathcal{R}}$ is of order $10^{-5}$). Clearly none of these give any indication of primordial non-gaussianity.

While the above results come from the CMB alone, other probes are becoming competitive (especially when considering more complex cosmological models), and so combination of data from different sources is of growing importance. We note that it has become fashionable to combine probes at the level of power-spectrum data vectors, taking into account nuisance parameters in each type of measurement. Recent examples include KiDS+GAMA [63] and Dark Energy Survey (DES) Year 1 [64]. For example, the DES analysis includes galaxy position–position clustering, galaxy–galaxy lensing, and weak lensing shear. Discussions on 'tension' in resulting cosmological parameters depend on the statistical approaches used. Commonly the cosmology community works within the Bayesian framework, and assesses agreement amongst data sets with respect to a model via Bayesian Evidence, essentially the denominator in Bayes's theorem. As an example of results, combining DES Y1 with *Planck*, BAO measurements from SDSS, 6dF, and BOSS, and type Ia supernovae from the Joint Lightcurve Analysis (JLA) dataset has shown the datasets to be mutually compatible and yields very tight constraints on cosmological parameters: $S_8 \equiv \sigma_8(\Omega_{\rm m}/0.3)^{0.5} = 0.799^{+0.014}_{-0.009}$, and $\Omega_{\rm m} = 0.301^{+0.006}_{-0.008}$ in $\Lambda$CDM, and $w = -1.00^{+0.04}_{-0.05}$ in $w$CDM [64]. The combined measurement of the Hubble constant within $\Lambda$CDM gives $H_0 = 68.2 \pm 0.6 \,{\rm km\,s^{-1}\,Mpc^{-1}}$, still leaving some level of tension with the local measurements described earlier. Future analyses and the next generation of surveys will test for deviations from $\Lambda$CDM, for example epoch-dependent $w(z)$ and modifications to General Relativity.

## 1.5.  Outlook for the future

The concordance model is now well established, and there seems little room left for any dramatic revision of this paradigm. A measure of the strength of that statement is how difficult it has proven to formulate convincing alternatives.

Should there indeed be no major revision of the current paradigm, we can expect future developments to take one of two directions. Either the existing parameter set will continue to prove sufficient to explain the data, with the parameters subject to ever-tightening constraints, or it will become necessary to deploy new parameters. The latter outcome would be very much the more interesting, offering a route towards understanding new physical processes relevant to the cosmological evolution. There are many possibilities on offer for striking discoveries, for example:

- the cosmological effects of a neutrino mass may be unambiguously detected, shedding light on fundamental neutrino properties;
- detection of primordial non-Gaussianities would indicate that non-linear processes influence the perturbation generation mechanism;

- detection of variation in the dark-energy density (*i.e.*, $w \neq -1$) would provide much-needed experimental input into its nature.

These provide more than enough motivation for continued efforts to test the cosmological model and improve its accuracy. Over the coming years, there are a wide range of new observations that will bring further precision to cosmological studies. Indeed, there are far too many for us to be able to mention them all here, and so we will just highlight a few areas.

The CMB observations will improve in several directions. A current frontier is the study of polarization, for which power spectrum measurements have now been made by several experiments. Detection of primordial *B*-mode anisotropies is the next major goal and a variety of projects are targeting this, though theory gives little guidance as to the likely signal level. Future CMB projects that are approved include *LiteBIRD* and the Simons Observatory.

An impressive array of cosmology surveys are already operational, under construction, or proposed, including the ground-based Hyper Suprime Camera (HSC) and Large Synoptic Survey Telescope (LSST) imaging surveys, spectroscopic surveys such as the Dark Energy Spectroscopic Instrument (DESI), and space missions *Euclid* and the Wide-Field Infrared Survey (WFIRST).

An exciting area for the future is radio surveys of the redshifted 21-cm line of hydrogen. Because of the intrinsic narrowness of this line, by tuning the bandpass the emission from narrow redshift slices of the Universe will be measured to extremely high redshift, probing the details of the reionization process at redshifts up to perhaps 20, as well as measuring large-scale features such as the BAOs. LOFAR and CHIME are the first instruments able to do this and have begun operations. In the longer term, the Square Kilometre Array (SKA) will take these studies to a precision level.

The development of the first precision cosmological model is a major achievement. However, it is important not to lose sight of the motivation for developing such a model, which is to understand the underlying physical processes at work governing the Universe's evolution. From that perspective, progress has been much less dramatic. For instance, there are many proposals for the nature of the dark matter, but no consensus as to which is correct. The nature of the dark energy remains a mystery. Even the baryon density, now measured to an accuracy of a percent, lacks an underlying theory able to predict it within orders of magnitude. Precision cosmology may have arrived, but at present many key questions remain to motivate and challenge the cosmology community.